\documentclass{aa}  

\usepackage{graphicx}
\usepackage{multirow}
\usepackage{caption}
\usepackage{multirow}
\usepackage{natbib}
\usepackage{hyperref}

\newcommand{\teff}{$T_{\rm eff}$}

\newcommand{\tef}{T_\mathrm{eff}}

\newcommand{\qq}{$\mathrm{q}^{2}$}

\bibpunct{(}{)}{;}{a}{}{,} 

\graphicspath{{plots/}}
\usepackage{txfonts}

\begin{document} 

 \title{Serendipitous discovery of the faint solar twin Inti 1\thanks{The data presented herein were obtained at the W.M. Keck Observatory, which is operated as a scientific partnership among the California Institute of Technology, the University of California, and the National Aeronautics and Space Administration. The Observatory was made possible by the generous financial support of the W.M. Keck Foundation.}}

   \author{Jhon Yana Galarza
          \inst{\ref{inst1}}
          \and
          Jorge Mel\'endez \inst{\ref{inst1}}
          \and
          Judith G. Cohen  \inst{\ref{inst2}}          }

  \institute{Universidade de S\~ao Paulo, IAG, Departamento de Astronomia, S\~ao Paulo, Rua do Mat\~ao 1226,  05508-090 SP, Brasil \\ \email{ramstojh@usp.br}\label{inst1}
  \and California Institute of Technology,  1200 E. California Blvd., MC 249-17, Pasadena, CA 91195, USA \label{inst2}
               }         
              
\titlerunning{Serendipitous discovery of Inti 1}


  \abstract
   {Solar twins are  increasingly the subject of many studies owing to their wide range of applications from testing stellar evolution models to the calibration of fundamental observables; these stars are also of interest  because high precision abundances  could be achieved that are key to investigating the chemical anomalies imprinted by planet formation. Furthermore, the advent of photometric surveys with large telescopes motivates the identification of faint solar twins in order to set the zero point of fundamental calibrations.}
   {We intend to perform a detailed line-by-line differential analysis to verify whether 2MASS J23263267-0239363 (designated here as Inti 1) is indeed a solar twin.}
   {We determine the atmospheric parameters and differential abundances using high-resolution ($R \approx 50 000$), high signal-to-noise (S/N $\approx$ 110 - 240 per pixel) Keck HIRES spectra for our solar twin candidate, the previously known solar twin HD 45184, and the Sun (using reflected light from the asteroid Vesta).}
   {For the bright solar twin HD 45184, we found $T_{\rm{eff}} = 5864 \pm 9$ K, log $g = 4.45 \pm 0.03$ dex, $v_{t} = 1.11 \pm 0.02$ $\rm{km\ {s}}^{-1}$, and [Fe/H]$ = 0.04 \pm 0.01$ dex, which are in good agreement with previous works. Our abundances are in excellent agreement with the recent high-precision work by \citet{Nissen:2015} with an element-to-element scatter of only 0.01 dex. The star Inti 1 has atmospheric parameters $T_{\rm{eff}} = 5837 \pm 11$ K, log $g = 4.42 \pm 0.03$ dex, $v_{t} = 1.04 \pm 0.02$ $\rm{km\ {s}}^{-1}$, and [Fe/H]$ = 0.07 \pm 0.01$ dex that are higher than solar. The age and mass of the solar twin HD 45184 (3 Gyr and 1.05 $\rm{M_{\odot}}$) and the faint solar twin Inti 1 (4 Gyr and 1.04 $\rm{M_{\odot}}$) were estimated  using isochrones. The differential analysis shows that HD 45184 presents an abundance pattern that is similar to typical nearby solar twins; this means this star has an enhanced refractory relative to volatile elements, while Inti 1 has an abundance pattern closer to solar, albeit somewhat enhanced in refractories. The abundance pattern of HD 45184 and Inti 1 could be reproduced by adding $\approx 3.5\ \rm{M_{\oplus}}$ and $\approx 1.5\ \rm{M_{\oplus}}$ of Earth-like material to the convective zone of the Sun.}
   {The star Inti 1 is a faint solar twin, therefore, it could be used to calibrate the zero points of different photometric systems. The distant solar twin Inti 1 has an abundance pattern similar to the Sun with only a minor enhancement in the refractory elements. It would be important to analyze other distant solar twins to verify whether they share the Sun's abundance pattern or if they are enhanced in refractories, as is the case in the majority of nearby solar twins.}


\maketitle
%

\section{Introduction}
\citet{Cayrel:1996} defines a solar twin as a star with the same atmospheric parameters (effective temperature, surface gravity, and microturbulent velocity) as the Sun. The first discovery of a solar twin was made by \citet{Porto:1997}, showing that 18 Sco has atmospheric parameters similar to the Sun. About a decade later three new solar twins were found \citep{King:2005, Melendez:2006, Takeda:2007}, however, these three solar twins and 18 Sco have an overabundance of lithium by a factor of 3-6 higher than the Sun. Later \citet{Melendez:2007} discovered HIP 56948, which is the best solar twin that we have until this point, with stellar parameters and abundances that are similar to the Sun and  a low lithium abundance \citep{Melendez:2012}.

In the last few years, the number of solar twins have increased to about 100 \citep{Pasquini:2008, Melendez:2009, Ramirez:2009, Takeda:2009, Baumann:2010, Onehag:2010, Datson:2012, Nascimento:2013, MelendezSchirbel:2014, Porto:2014, Ramirez:2014-1}. Solar twins could be a useful source for astronomical tests and applications. For example, they could set the zero point of fundamental calibrations \citep{Holmberg:2006, Casagrande:2010, Datson:2014},  and they are used to subtract the Sun's reflected light on asteroids to study their mineralogy \citep[e.g.,][]{Lazzaro:2004, Jasmim:2013}. Solar twins  are also useful for testing stellar interior and evolution models \citep[e.g.,][]{Tucci:2015},  investigating the chemical evolution of the Galactic disk \citep{Nissen:2015, Spina:2015}, studying the rotational evolution of the Sun \citep{Nascimento:2013, Nascimento:2014}, and measuring distances using spectroscopically identified solar twins \citep{Jofree:2015}. 

Another application of solar twins is the study of refractories locked by planet formation. \citet{Melendez:2009} found that the difference between abundances of the Sun relative to the solar twins is not zero, showing that the Sun presents abundance anomalies that are correlated with the condensation temperature, $T_{C}$, \citep{Lodders:2003}. The Sun presents a deficiency in refractory ($T_{C} \leq 900$ K) relative to volatile ($T_{C} \geq 900$ K) elements, which could be a signature of rocky planet formation \citep{Chambers:2010}. Later, \citet{Ramirez:2009} supported this idea by studying the abundance of 64 stars with fundamental parameters similar to the Sun; their result showed a strong relation between the abundances and condensation temperature in the majority of their sample. Other authors have been investigating this relation with the existence of planets, although the results are still not conclusive \citep{Gonzales:2010, Schuler:2011, Adibekyan:2014, Maldonado:2015, Nissen:2015}. Are the abundance anomalies a phenomenon restricted to the solar neighborhood or is it also present in other regions of the Galaxy? This could be investigated using distant solar twins.

In this paper, we report the discovery of the faint solar twin 2MASS J23263267-0239363, which we call  Inti 1\footnote{Inti is Quechua for `Sun'} for practical reasons.

\section{Spectroscopy observations and data reduction}
\begin{figure}
\centering
\includegraphics[width=9cm]{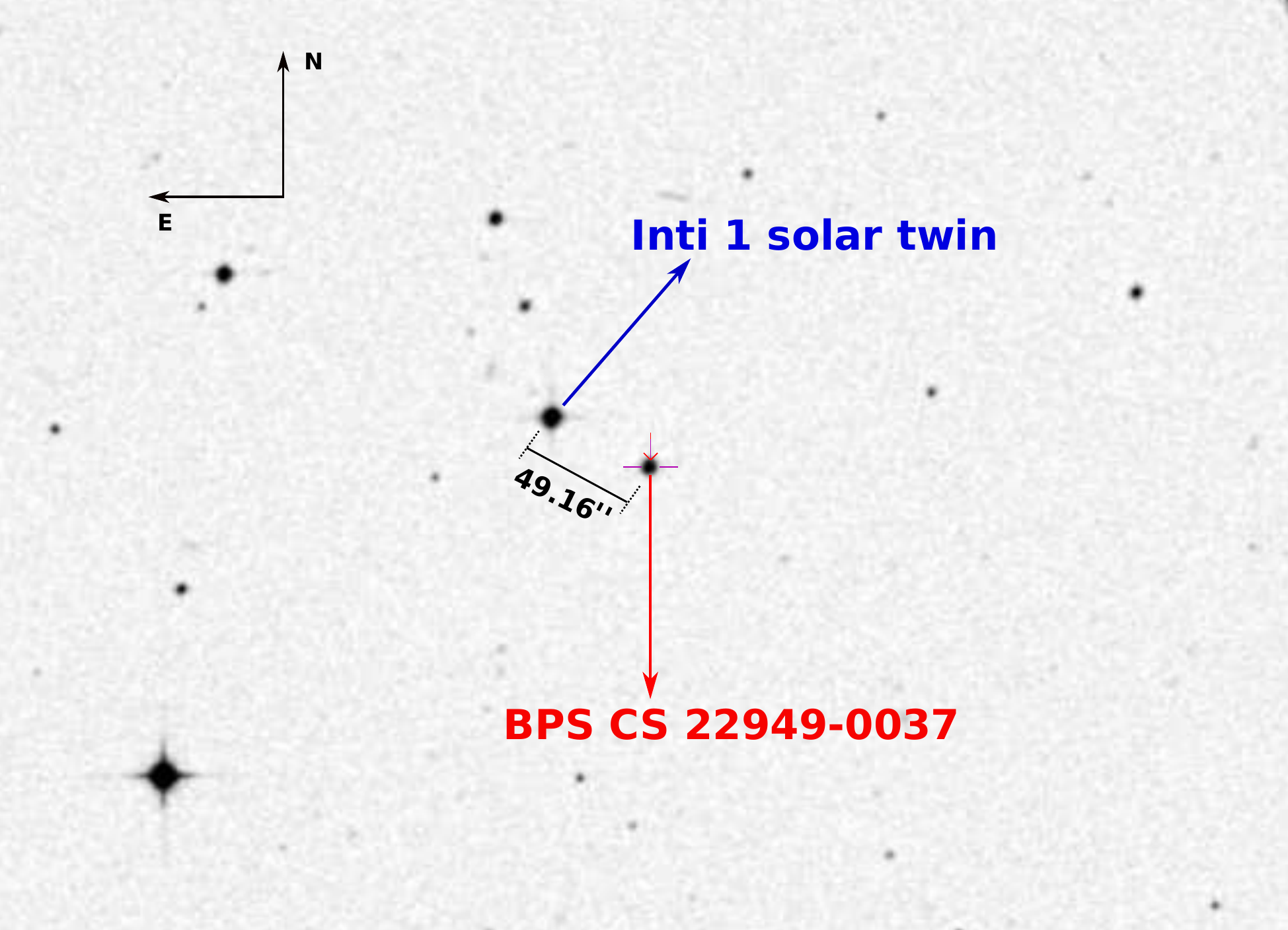}
\caption{Separation between Inti 1 and BPS CS 22949-0037 is 49.16''. Image from the Aladin Sky Atlas (\object{http://aladin.ustrasbg.fr}).}
\label{fig:2mass}
\end{figure}

\begin{figure*}
\centering
\includegraphics[width=18cm]{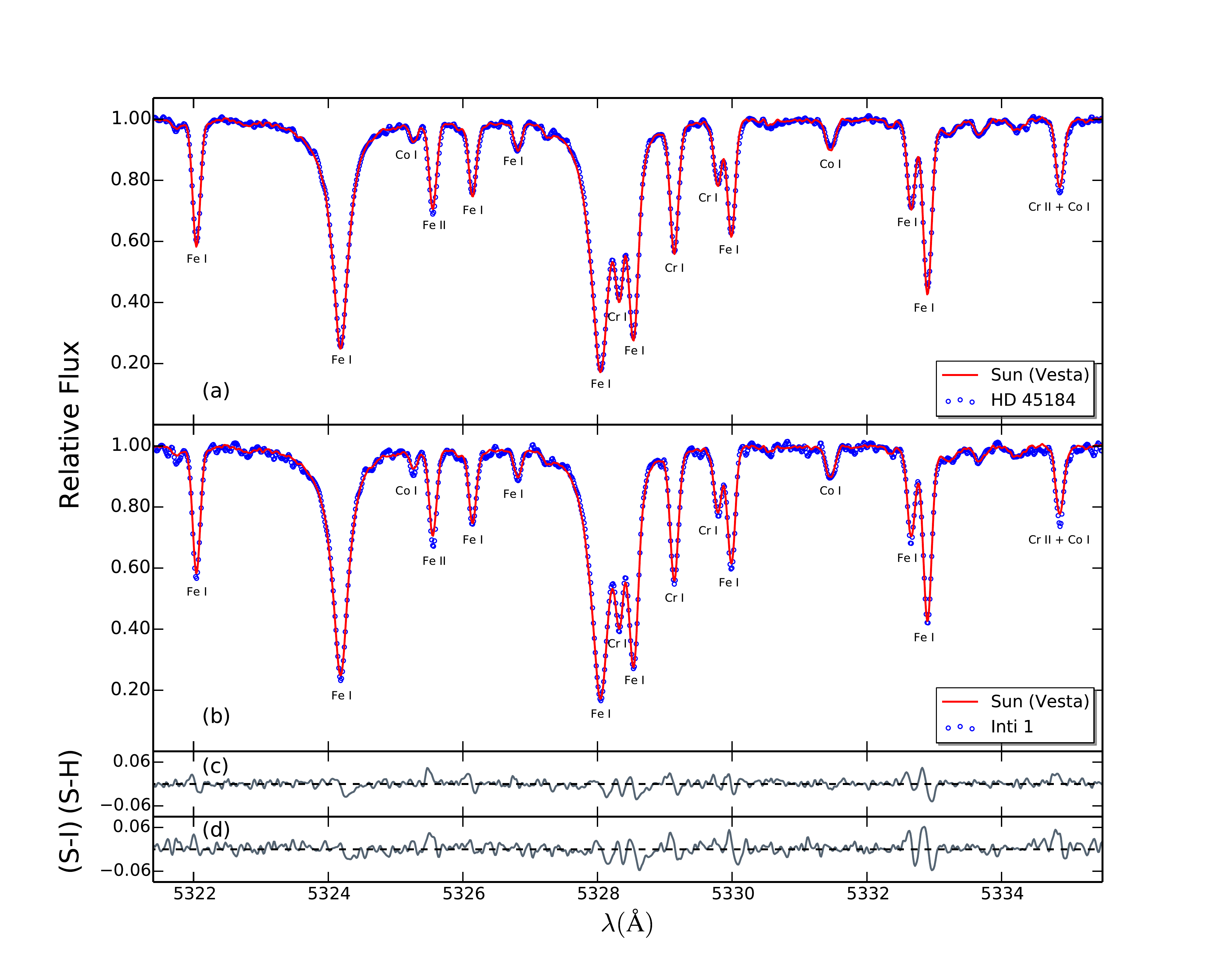}
\caption{Comparison between the spectra of HD 45184, Inti 1, and the Sun in the region 5321-5336 $\AA$. Panel (a) shows the spectra of HD 45184 (blue open circles) and the Sun (red solid line). Panel (b) shows the spectra of the Sun (red solid line) and candidate solar twin Inti 1 (blue open circles), showing a striking similarity. The different chemical composition is revealed through careful line- by-line measurements. Residuals between the Sun and HD 45184 (S-H), and the Sun and Inti 1 (S-I) are shown in the panel (c) and (d), respectively.}
\label{fig:stellar-espectra}
\end{figure*}

Inti 1 was observed with the HIRES spectrograph \citep{Vogt:1994} at the Keck I telescope at the coordinates ($\alpha$ = 23:26:32.61, $\delta$ = -2:39:35.3) because it was mistaken for the carbon star BPS CS 22949-0037, whose coordinates are actually ($\alpha$ = 23:26:29.80, $\delta$ = -2:39:57.94). This mistake was due to their proximity (Fig. \ref{fig:2mass}). We intended to take two exposures of 1200 s, but after the first exposure it was realized that the star observed was not BPS CS 22949-0037, and the planned exposure sequence was terminated after only one exposure.
Fortunately, in the same night we observed the Sun using the same setup. A recent inspection of the spectrum of the wrongly observed target revealed a similarity to the solar spectrum (see Fig. \ref{fig:stellar-espectra}), thus making Inti 1 a good candidate to be a solar twin, which we confirmed after high precision analysis. During the same observing run, we also obtained a spectrum for the bright solar twin HD 45184 \citep[e.g.,][]{Nissen:2015}.

The spectra of HD 45184, Inti 1 and the Sun (using solar reflected light from the Vesta asteroid), were obtained with HIRES/Keck on November 1-3, 2004, covering the wavelength region from 3190-5980 $\AA$. The exposure time for HD 45184, Inti 1 and Vesta  were 2$\times$300 s, 1200 s, and 100 s. The spectral resolution, $R=\lambda / \lambda \Delta$ is about 50000, while the signal-to-noise ratios (S/N) estimated for HD 45184, Inti 1 and Vesta are 160, 110 and 240 per pixel.

The spectra were extracted using the MAKEE\footnote{\object{www.astro.caltech.edu/$\sim$tb/}} pipeline that was specifically designed by T. Barlow to reduce HIRES spectra. The standard procedure is followed with bias subtraction, flat fielding, sky subtraction, order extraction, and wavelength calibration. We note that MAKEE already applies the heliocentric correction, so that the output wavelengths are heliocentric. Further processing was performed with IRAF\footnote{IRAF is distributed by the National Optical Astronomy Observatory, which is operated by the Association of the Universities for Research in Astronomy, Inc. (AURA) under cooperative agreement with the National Science Foundation.}.

We estimated the radial velocity of the star using the {\tt rvidlines} task in IRAF, and then the  spectrum was corrected to the rest frame with the {\tt dopcor} task. The spectra were normalized using IRAF's {\tt continuum} task with orders of the polynomial {\tt spline} varying from 1 to 5 for each order. In Fig. \ref{fig:stellar-espectra} we show a part of the reduced spectra of HD 45184, Inti 1, and the Sun in the region 5321-5336 $\AA$, where we can see that both stars have an impressive resemblance to the Sun.

\begin{figure*}
\centering
\includegraphics[width=19cm]{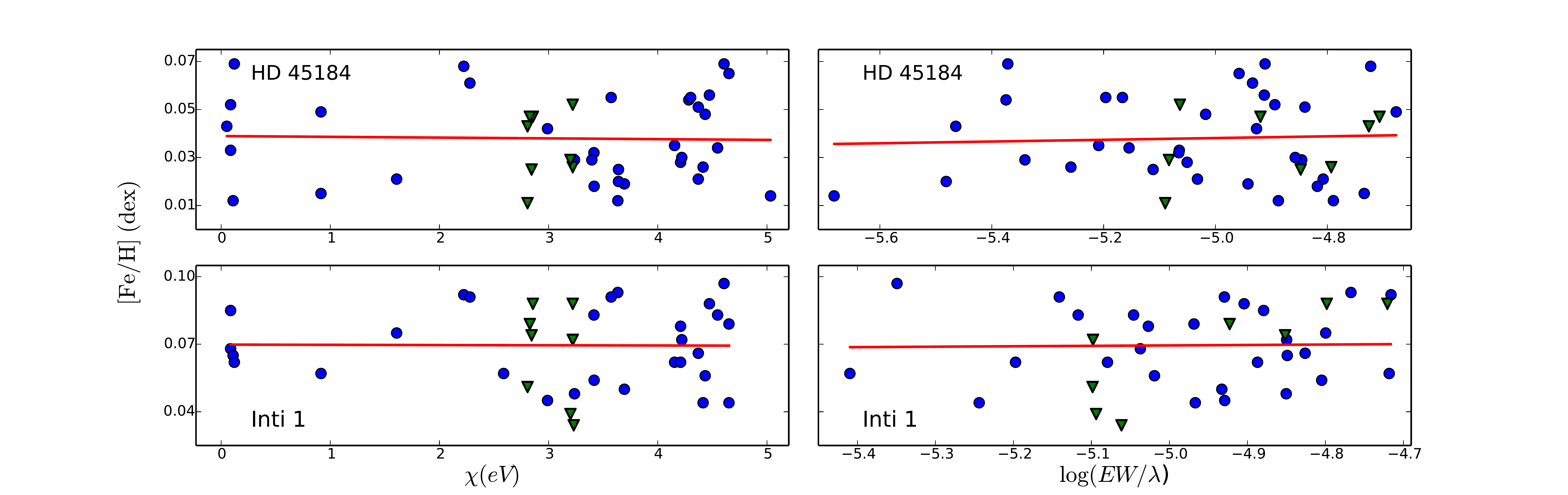}
\caption{Differential iron abundances as a function of excitation potential (left panels) and reduced equivalent width (right panels) for the solar twin HD 45184 and Inti 1. Blue filled circles and green filled triangles represent Fe I and Fe II, while red lines are the fits.}
\label{fig:stellar-abundance}
\end{figure*}

The similarity between Inti 1 and the Sun also extends to  colors; using VIZIER\footnote{\object{http://vizier.u-strasbg.fr}}, we found the magnitudes\footnote{$\mathrm{V, I} $ magnitudes adopted from APASS (The AAVSO Photometric All-Sky Survey), $\mathrm{J, H, K}$ magnitudes adopted from \citet{Cutri:2003}, $\mathrm{W1, W2}$ magnitudes adopted from \citet{Cutri:2012}.}: $\mathrm{V} = 12.857 \pm 0.028$, $\mathrm{B} = 13.516 \pm 0.032$, $\mathrm{I} = 12.045 \pm 0.072$, $\mathrm{J} = 11.559 \pm 0.023$, $\mathrm{H} = 11.247 \pm 0.023$, $\mathrm{K} = 11.168 \pm 0.024$, $\mathrm{W1} = 11.125 \pm 0.023$, and $\mathrm{W2} = 11.180 \pm 0.022$. Inti 1 and the Sun present similar color magnitudes (Table \ref{tab:magnitude}) reinforcing the idea that it is a solar twin. Extinction coefficients for $\mathrm{(B - V)}$, $\mathrm{(V - I)}$, $\mathrm{(V - J)}$, $\mathrm{(V - H)}$, $\mathrm{(V - K)}$ photometry were taken from \citet{Ramirez:2005-2} and for $\mathrm{(V - W1)}$, $\mathrm{(V - W2)}$ from \citet{Yuam:2013}. For the extinction correction, we used $\mathrm{E(B - V)} \footnote{\object{http://irsa.ipac.caltech.edu/frontpage}} = 0.044 \pm 0.002$ , that is from the reddening maps of \citet{Schlegel:1998}, with the correction proposed by \citet{Schlafly:2011}.

\begin{table}
\caption{Comparison of colors of the Sun and Inti 1.}
\label{tab:magnitude}
\centering 
\renewcommand{\footnoterule}{}  
\begin{tabular}{cccc} 
\hline    
\hline 
 \multirow{2}{*}{Color}   & \multicolumn{2}{c}{Inti 1}                 & \multirow{2}{*}{Sun} \\ 
                          &    Observed         &   Derredened         \\\hline
$\mathrm{(B - V)}^{a}$  & $0.659 \pm 0.042$ & $0.615 \pm 0.042$  & $0.653 \pm 0.005$\\  
$\mathrm{(V - I)}^{a}$  & $0.812 \pm 0.077$ & $0.755 \pm 0.077$  & $0.702 \pm 0.010$\\ 
$\mathrm{(V - J)}^{b}$  & $1.298 \pm 0.036$ & $1.203 \pm 0.036$  & $1.198 \pm 0.005$\\  
$\mathrm{(V - H)}^{b}$  & $1.610 \pm 0.036$ & $1.500 \pm 0.036$  & $1.484 \pm 0.009$\\  
$\mathrm{(V - K)}^{b}$  & $1.689 \pm 0.036$ & $1.570 \pm 0.036$  & $1.560 \pm 0.008$\\  
$\mathrm{(V - W1)}^{b}$ & $1.731 \pm 0.036$ & $1.603 \pm 0.036$  & $1.608 \pm 0.008$\\  
$\mathrm{(V - W2)}^{b}$ & $1.676 \pm 0.035$ & $1.556 \pm 0.035$  & $1.583 \pm 0.008$\\  
\hline       
\end{tabular}
\tablefoot{\\
a : predicted solar colors from \cite{Ramirez:2012} \\
b : predicted solar colors from \cite{Casagrande:2012}
}
\end{table}


\section{Abundance analysis}
\label{sec:3}
We adopted the line list of \cite{MelendezIvanAmanda:2014} to measure the equivalent widths (EWs) of the spectral lines; we used the IRAF {\tt{splot}} task and fitted the line profiles using Gaussians. The pseudo-continuum regions were determined as in \cite{Bedell:2014} in a window of 8 $\AA$.

\begin{table*}
\caption{Comparison of stellar parameters for the solar twin HD 45184}
\label{tab:sources}
\centering 
\renewcommand{\footnoterule}{}  
\begin{tabular}{ccccccccccccl} 
\hline    
\hline 
 $T_{\rm{eff}}$  & error   &  log $g$ & error & $\rm{[Fe/H]}$& error & $v_{t}$ & error & Age & error & Mass & error&  Source  \\
    (K)     &   (K)   &   (dex)  & (dex) & (dex)        & (dex) &  (km\ $\rm{s}^{-1}$)       &   (km\ $\rm{s}^{-1}$) & (Gyr) & (Gyr) & $\rm{M}_{\odot}$ & $\rm{M}_{\odot}$ &        \\
\hline
    5864    &   9     &  4.45  &  0.03       &   0.040       & 0.010       &  1.11   & 0.02        & 3.0 & 1.2 & 1.05 & 0.01 & This work  \\
    5873    &   18    &  4.41  &  0.04       &   0.070       & 0.016       &  1.03   & 0.04        & 3.7 & 1.2 & 1.06 & 0.02 & 1$\dagger$  \\
    5871    &   6     &  4.45  &  0.01       &   0.047       & 0.006       &  1.06   & 0.02        & 2.7 & 0.5 & 1.06 & 0.01 & 2$\dagger$  \\ 
    5833    &   10    &  4.37  &  0.02       &   0.010       & 0.010       &  1.04   & 0.06        & ... & ... & ...  & ...  & 3$\dagger$  \\ 
    5849    &   86    &  4.45  &  0.11       &   0.040       & 0.090       &  1.11   & 0.09        & 4.4 & 2.3 & 1.03 & 0.05 & 4$\dagger$  \\ 
    5869    &   14    &  4.47  &  0.02       &   0.040       & 0.010       &  1.03   & 0.04        & 2.3 & ... & 1.05 & ...  & 5*$\dagger$ \\ 
    5863    &   5     &  4.44  &  0.01       &   0.040       & 0.004       &  1.05   & 0.02        & 2.9 & 0.5 & 1.06 & 0.01 & 6  \\ 
\hline       
\hline
\end{tabular}
\tablefoot{\\ 1) \citet{Spina:2015}; 2) \citet{Nissen:2015}; 3) \citet{Maldonado:2015A}; 4) \citet{Bensby:2014}; 5) \citep{Sousa:2008,Delgado:2014}; 6) weighted mean from the literature.
\\ * Stellar parameters from \citet{Sousa:2008}; age and mass from \citet{Delgado:2014} and \citet{Schneider:2011}. \\ $\dagger$ The parameters reported here are based on the differential stellar parameters and adopting for the Sun \teff = 5777 K, log $g = 4.44$ dex and $v_{t} = 1.00\ $km$\ \rm{s}^{-1}$.}
\end{table*}

In order to compute the stellar parameters, we employed the MARCS model atmospheres \citep{Gustafsson:2008} and the 2014 version of the local thermodynamic equilibrium (LTE) code MOOG \citep{Sneden:1973}. The LTE is adequate for differential studies of solar twins, as the differential NLTE (non-LTE) effects in solar twins are negligible \citep{Melendez:2012, Monroe:2013}.

Spectroscopic equilibrium is used to determine the effective temperature ($T_{\rm{eff}}$), surface gravity (log $g$), metallicity ([Fe/H]), and microturbulent velocity ($v_{t}$). This is performed by imposing the relative excitation and ionization equilibrium, and evaluating the lack of dependence of differential iron abundance with reduced equivalent width ($\rm{EW} / \lambda$). All the calculations are strictly differential ($\Delta A_{i}$)\footnote{The differential abundances is defined as $\Delta A_{i} = A^{*}_{i} - A^{\odot}_{i}$ }.

As a first step, we determined the abundances of the Sun ($A^{\odot}_{i}$) for each line, adopting the standard solar parameters ($T_{\rm{eff}}=5777$ K, log $g = 4.44$ dex \citep{Cox:2000}, and $v_{t} = 1.00\ $km$\ \rm{s}^{-1}$ as in \citet{Ramirez:2014-1}). The second step is the determination of stellar parameters of Inti 1; initially we set as the atmospheric parameters those of the Sun, then the abundances ($A^{\star}_{i}$) were found running the MOOG code. We achieved the spectroscopic solution varying the values for the parameters until achieving differential spectroscopic equilibrium (Fig. \ref{fig:stellar-abundance}). 

\begin{figure}
\centering
\includegraphics[width=9cm]{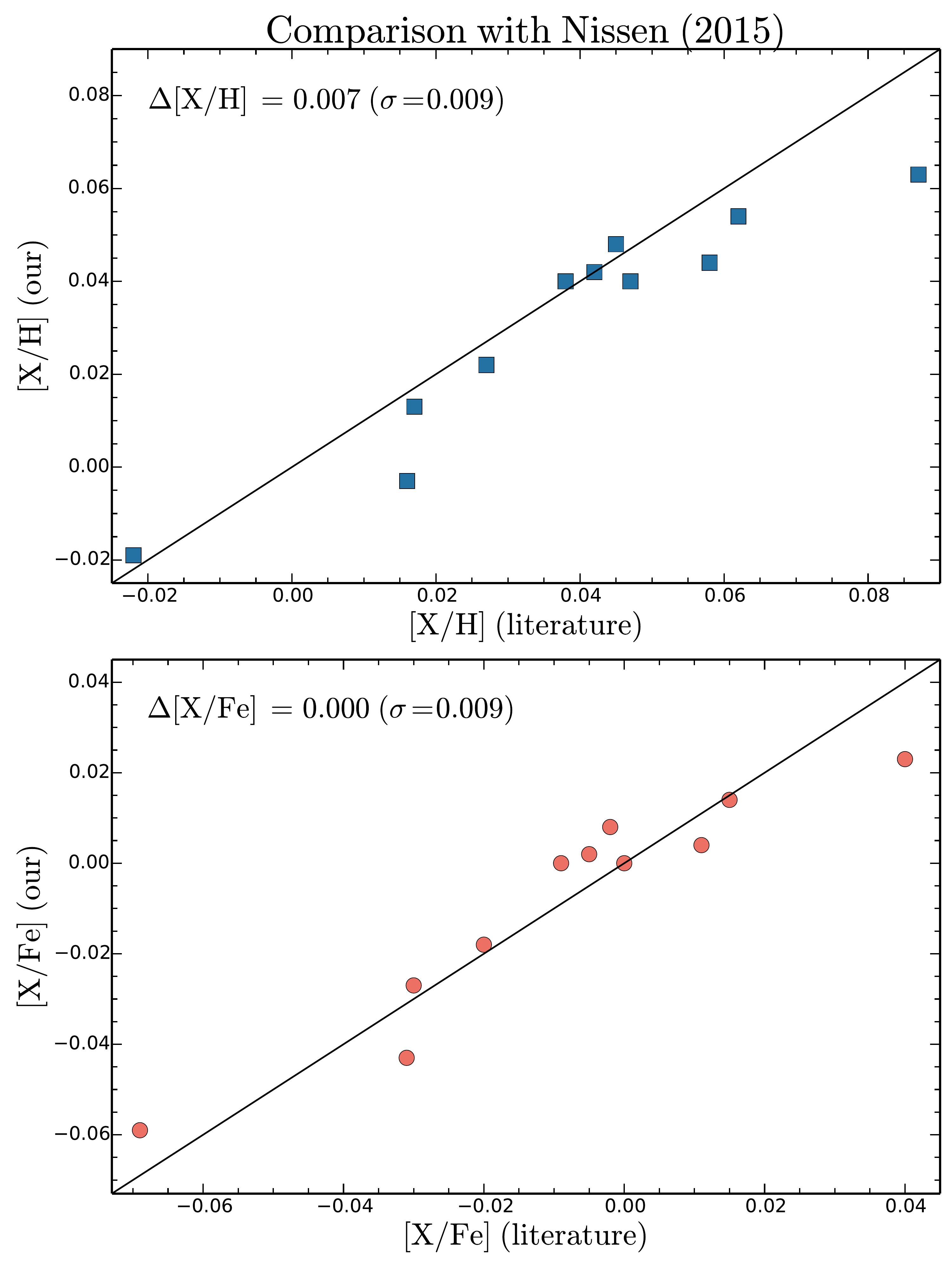}
\caption{Upper panel: Comparison between our [X/H] ratios and those from the high precision work by \cite{Nissen:2015}. Lower panel: As above for [X/Fe].}
\label{fig:comparison}
\end{figure}

We first study the bright solar twin HD 45184 to validate our method. The stellar parameters that we found are $T_{\rm{eff}} = 5864 \pm 9$ K, log $g = 4.45 \pm 0.03$ dex, $v_{t} = 1.11 \pm 0.02$ $\rm{km\ {s}}^{-1}$, and [Fe/H]$\ = 0.04 \pm 0.01$ dex. These values are in a good agreement with all previous works found in the literature, as shown in Table \ref{tab:sources}, as well as in good accord with the weighted mean value, $T_{\rm{eff}} = 5863 \pm 5$ K, log $g = 4.44 \pm 0.01$ dex, $v_{t} = 1.05 \pm 0.02$ $\rm{km\ {s}}^{-1}$, and [Fe/H]$\ = 0.040 \pm 0.004$ dex. The stellar parameters found for Inti 1 are $T_{\rm{eff}}= 5837 \pm 11$ K, log $g = 4.42 \pm 0.03$ dex, $v_{t} = 1.04 \pm 0.02$ km\ $\rm{s}^{-1}$, and [Fe/H]$\ = 0.07 \pm 0.01$ dex.  According to the solar twin definition of \citet{Ramirez:2009}, a solar twin should be within $\Delta T_{\rm{eff}} = 100$ K,  $\Delta$log$\ g = 0.1$ dex, and $\Delta$[Fe/H] = 0.1 dex of the Sun; our results agree with this definition, hence, Inti 1 is a solar twin.

Once the atmospheric parameters were set, we measured the abundances of 18 elements other than iron using atomic lines: C, Na, Mg, Si, Ca, Sc, Ti, Cr, Mn, Co, Ni, Cu, Zn, Y, Zr, Ce, and Nd. For carbon we also used CH molecular lines and for nitrogen we used NH. The hyperfine correction was made for Mn, Co, Y, and Cu, adopting the HFS data from \citet{MelendezIvanAmanda:2014}. We compare our abundances for the bright solar twin HD 45184 with the most precise abundances available (Fig. \ref{fig:comparison}), which are from \citet{Nissen:2015}. We found a difference (this work - \citet{Nissen:2015}) of $\Delta$[X/H] = 0.007 ($\sigma$ = 0.009 dex) and $\Delta$[X/Fe] = 0.000 ($\sigma$ = 0.009 dex). This shows that our precision is about 0.01 dex. We also compared with two other works in the literature, resulting in $\Delta$[X/H] = 0.023 ($\sigma$ = 0.016 dex) and $\Delta$[X/Fe] = - 0.007 ($\sigma$ = 0.016 dex) for \citet{Spina:2015}, and $\Delta$[X/H] = 0.014 ($\sigma$ = 0.030 dex) and $\Delta$[X/Fe] = 0.004 ($\sigma$ = 0.030 dex) for \citet{Gonzales_Hernandez:2010}.

\begin{figure}
\centering
$\begin{array}{c}
\includegraphics[width=9cm]{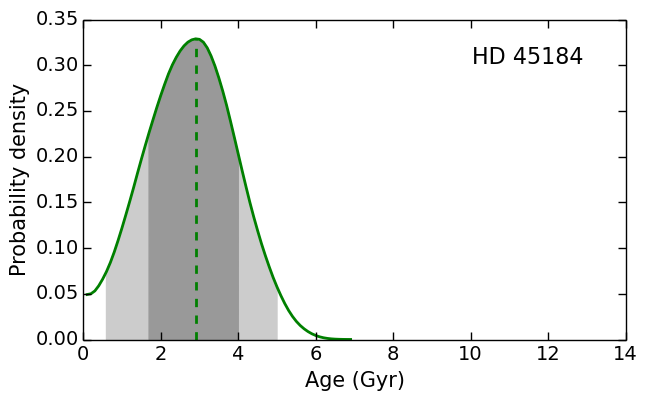}\\
\includegraphics[width=9cm]{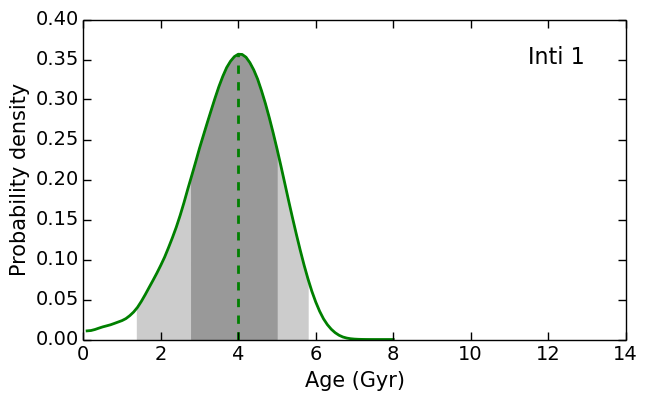}
\end{array}$
\caption{Age probability distribution of HD 45184 (upper panel) and Inti 1 (lower panel). The dashed lines represent the most probable age. The different regions represent $\pm$ 1 sigma and $\pm$ 2 sigma confidence level.}
\label{fig:mass-age}
\end{figure}

The differential abundances for HD 45184 and Inti 1 are presented in Table \ref{tab:adundances-hd}, including the observational errors and systematic errors (due to uncertainties in the stellar parameters), and we also included the total error obtained from quadratically adding the statistical and systematic errors.

\begin{figure}
\centering
\includegraphics[width=9cm]{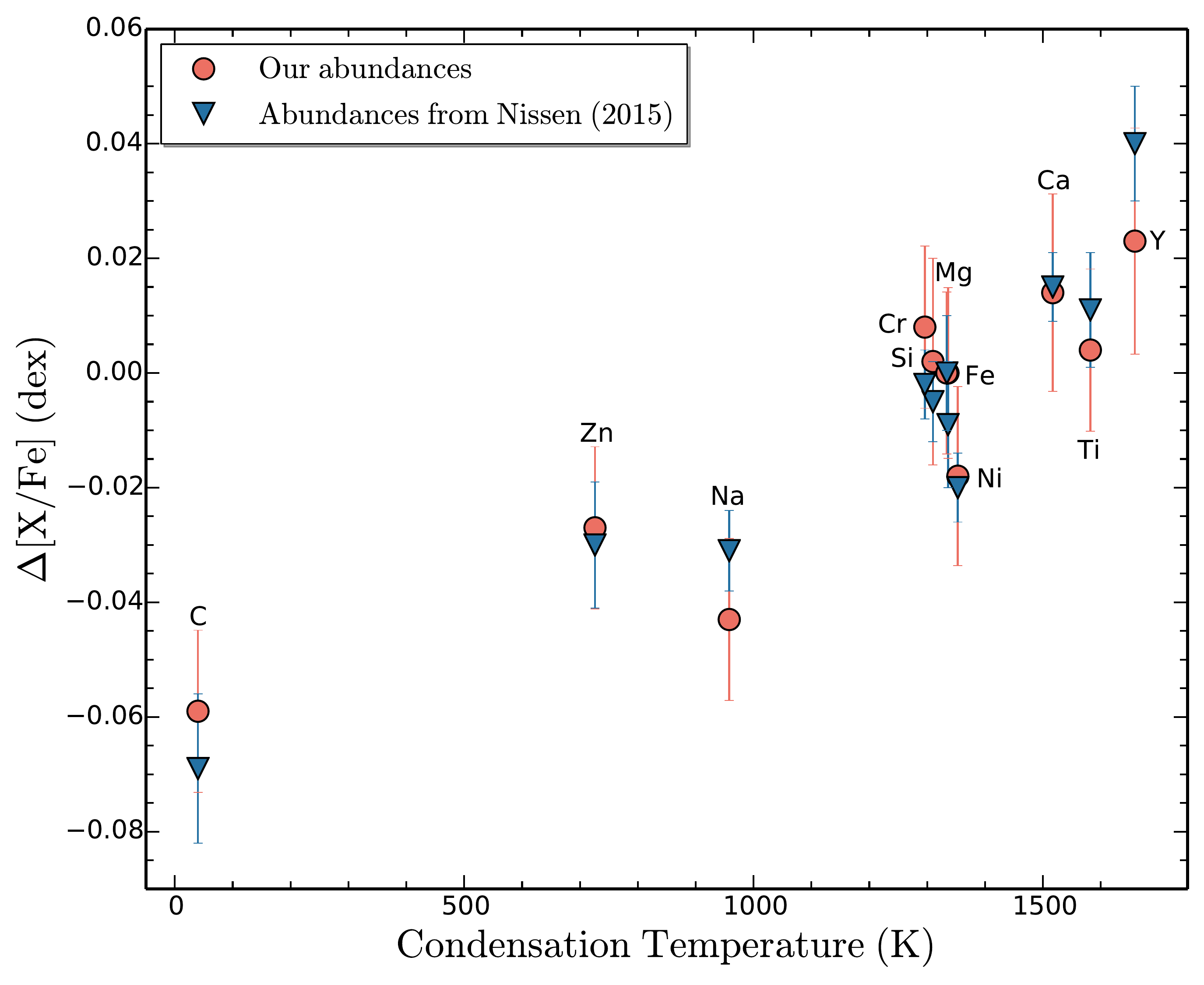}
\caption{Comparison between our abundance and \citet{Nissen:2015}, as a function of condensation temperature \citep{Lodders:2003}.}
\label{fig:TcAB_comparison}
\end{figure}

\begin{figure*}
\centering
\includegraphics[width=19.cm]{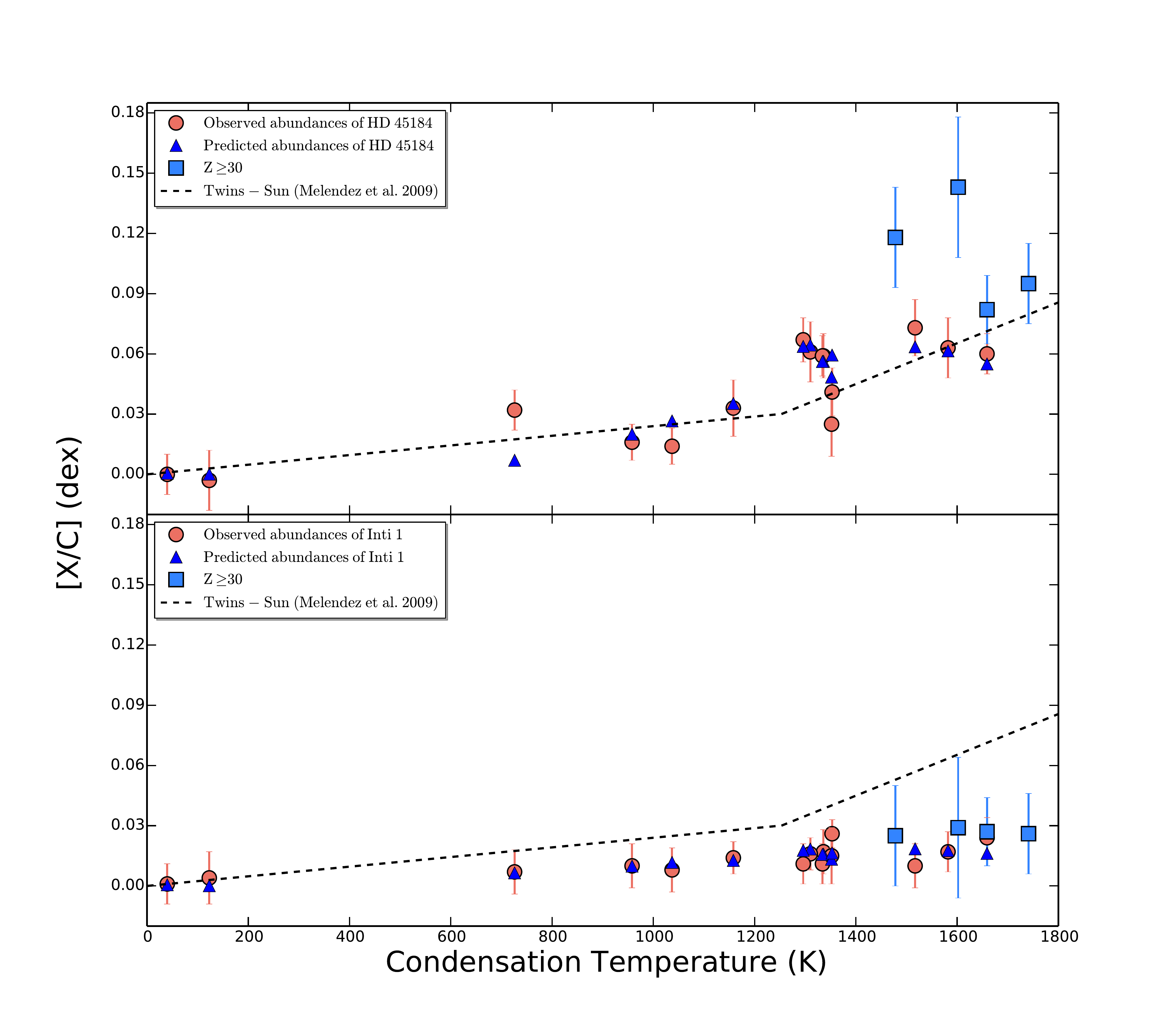}
\caption{Upper panel: Abundance pattern of HD 45184 (red filled circles) versus condensation temperature. We reproduce the chemical pattern of HD 45184 (blue filled triangles) adding $3.5\ \mathrm{M_{\oplus}}$ of Earth-like material to the convective zone of the Sun. Lower panel: The differential abundance ratios of Inti 1 (red filled circles) relative to the Sun as a function of condensation temperature. The mean abundance pattern of 11 solar twins studied by \cite{Melendez:2009} is represented by the black dashed line. We could reproduce the trend of Inti 1 (blue filled triangles), adding $1.5\ \mathrm{M_{\oplus}}$ of Earth-like material to the convective zone of the Sun. Neutron capture elements are represented by squares in both panels.}
\label{fig:abundance-tc}
\end{figure*}

We computed the mass and age of HD 45184 and Inti 1 using our atmospheric parameters and employing the $\mathrm{q}^{2}$ (qoyllur-quipu) code\footnote{\object{https://github.com/astroChasqui/q2}} . This code is based in python and was developed by \citet{Ramirez:2014-1}, to determine the masses and ages of stars using the Yonsei-Yale isochrones \citep[e.g.,][]{Yi:2001}. 

Our results show a mass of 1.05 $\rm{M_{\odot}}$ and 1.04 $\rm{M_{\odot}}$ for HD 45184 and Inti 1. The age obtained is about 3 Gyr and 4 Gyr for HD 45184 and Inti 1 (Fig. \ref{fig:mass-age}). The mass and age of HD 45184 are in excellent agreement with the weighted mean value found in the literature (Table \ref{tab:sources}). The Yonsei-Yale grids of stellar models do not include nonstandard physics such as those included in studies of lithium depletion \citep{Nascimento:2009, Denissenkov:2010, Li:2012}. We tested our method for the star 18 Sco, which was analyzed using nonstandard models by \citet{Li:2012}, and found the same mass and age. Furthermore, \citet{MelendezIvanAmanda:2014} shows that the exact choice of isochrone do not have an important impact on the masses and ages relative to the Sun.

\section{Abundance trends and results}
\label{sec:4}
A comparison between our chemical pattern and that of \citet{Nissen:2015} is shown in Fig. \ref{fig:TcAB_comparison}. It is reassuring that we get the same trend as the precise work by \citet{Nissen:2015}. In Fig. \ref{fig:abundance-tc} we show the differential abundances normalized to carbon for HD 45184 (upper panel) and Inti 1 (lower panel), as a function of condensation temperature \citep{Lodders:2003}. The chemical pattern of HD 46184 follows the abundance pattern of the fit of the mean trend of 11 solar twins studied by \citet{Melendez:2009}, as shown by the dashed lines after a vertical shift. Inti 1 has an abundance pattern closer to solar, albeit slightly enhanced in refractories. The offset between the refractory and volatiles elements of HD 45184 and Inti 1 are 0.076 dex and 0.025 dex, respectively. 

\begin{table}
\caption{Fundamental parameters for Inti 1}
\label{tab:results}
\centering 
\renewcommand{\footnoterule}{}  
\begin{tabular}{cc} 
\hline    
\hline 
      Inti 1          &    Parameters    \\ 
\hline
    $T_{\rm{eff}}$    & $ 5837 \pm 11$ K  \\  
  log $g$             & $4.42 \pm 0.03$ dex    \\
$\rm[Fe/H]$           & $0.07 \pm 0.01$ dex    \\ 
$v_{t}$               & $1.04 \pm 0.02 $ km\ $\rm{s}^{-1}$   \\
Mass                  & $1.04^{+0.01}_{-0.08}$ $\ \mathrm{M_{\odot}}$ **\\
Distance              & $389.70 \pm 36.00 $ pc *          \\
Convective Mass       & $0.019$ $\ \mathrm{M_{\odot}}$ ***  \\
Age                   & $4.00^{+1.00}_{-1.26}$ Gyr **   \\ 
logL                  & $0.06 \pm 0.032 \ \mathrm{L_{\odot}}$ **     \\
$\mathrm{M_{V}}$      & $4.68^{+0.08}_{-0.07}$ mag **           \\
Radius                & $1.05 \pm 0.04 \ \mathrm{R_{\odot}}$ **   \\
\hline       
\end{tabular}
\tablefoot{\\
 Error of the distance modulus was calculated assuming an uncertainty of 0.2 mags. \\
** From the q2 code.  \\
*** Using \citet{Siess:2000} models.
}
\end{table}

\citet{Nissen:2015} showed that the [Y/Mg] ratio is an age indicator. Using his relation, we estimate ages of 3.7 $\pm$ 0.6 Gyr and 4.1 $\pm$ 0.6 Gyr for HD 45184 and Inti 1; these results are in agreement with the isochronal ages computed by using \qq. 

Following \cite{Chambers:2010}, we computed the mass of rocky material needed to reproduce the abundance trend of HD 45184 and Inti 1. Our results show a value of $3.5 \pm 1\ \rm{M_{\oplus}}$ of refractory elements using a mixture of Earth-like and CM-chondrite-like material for HD 45184 and $1.5 \pm 1\ \rm{M_{\oplus}}$ for Inti 1. These patterns are represented by blue filled triangles in Fig. (\ref{fig:abundance-tc}). 

\begin{table*}
\caption{Stellar abundances [X/H] and errors for HD 45184 and Inti 1.}
\label{tab:adundances-hd}
\centering 
\renewcommand{\footnoterule}{}  
\begin{tabular}{lrrrrrrrrrr} 
\hline    
\hline
         &        &       &       & HD 45184&       &       &       &    \\
\hline 
\hline
{Element}& $\Delta$ [X/H]   & $\Delta \tef$ & $\Delta$log $g$ & $\Delta v_t$      & $\Delta$[Fe/H]  & Param\tablefootmark{a} & Obs\tablefootmark{b} & Total\tablefootmark{c}  \\
{}       & LTE              &   \textbf{$\pm$}9 K        &  \textbf{$\pm$}0.03 dex      & \textbf{$\pm$}0.02 km\ s$^{-1}$ & \textbf{$\pm$}0.01 dex       &                        &                      &       \\
{}       & (dex)            & (dex)         & (dex)           & (dex)             & (dex)           & (dex)                  & (dex)                & dex   \\
\hline
C        & -0.019 & 0.005 & 0.007 & 0.000 & 0.000 & 0.009 & 0.006 & 0.010 \\
Na       & -0.003 & 0.004 & 0.002 & 0.000 & 0.000 & 0.004 & 0.008 & 0.009 \\
Mg       &  0.040 & 0.007 & 0.003 & 0.004 & 0.000 & 0.009 & 0.007 & 0.011 \\
Si       &  0.042 & 0.002 & 0.002 & 0.000 & 0.001 & 0.003 & 0.001 & 0.015 \\
Ca       &  0.054 & 0.005 & 0.003 & 0.004 & 0.000 & 0.007 & 0.012 & 0.014 \\
Sc       &  0.041 & 0.007 & 0.003 & 0.000 & 0.000 & 0.008 & 0.006 & 0.010 \\
Ti       &  0.044 & 0.009 & 0.002 & 0.003 & 0.000 & 0.010 & 0.012 & 0.015 \\
Cr       &  0.048 & 0.007 & 0.001 & 0.004 & 0.000 & 0.008 & 0.008 & 0.011 \\
Mn       &  0.014 & 0.006 & 0.001 & 0.003 & 0.000 & 0.007 & 0.012 & 0.014 \\
Fe       &  0.040 & 0.007 & 0.001 & 0.006 & 0.000 & 0.010 & 0.003 & 0.010 \\
Co       &  0.006 & 0.006 & 0.003 & 0.001 & 0.000 & 0.007 & 0.014 & 0.016 \\
Ni       &  0.022 & 0.004 & 0.001 & 0.003 & 0.001 & 0.005 & 0.011 & 0.012 \\
Cu       & -0.005 & 0.004 & 0.002 & 0.002 & 0.001 & 0.005 & 0.007 & 0.009 \\
Zn       &  0.013 & 0.002 & 0.001 & 0.007 & 0.002 & 0.008 & 0.006 & 0.010 \\
Y        &  0.063 & 0.001 & 0.010 & 0.009 & 0.002 & 0.014 & 0.010 & 0.017 \\
Zr       &  0.076 & 0.001 & 0.013 & 0.004 & 0.002 & 0.015 & 0.014 & 0.020 \\
Ce       &  0.099 & 0.002 & 0.014 & 0.002 & 0.003 & 0.015 & 0.020 & 0.025 \\
Nd       &  0.124 & 0.002 & 0.014 & 0.001 & 0.003 & 0.014 & 0.032 & 0.035 \\
C(CH)    & -0.021 & 0.007 & 0.001 & 0.001 & 0.006 & 0.009 & 0.013 & 0.016 \\
N(NH)    &  0.003 & 0.009 & 0.001 & 0.000 & 0.006 & 0.011 & 0.011 & 0.015 \\

\hline 
\hline
         &        &       &       & Inti 1&       &       &       &    \\\hline
\hline
{Element}& $\Delta$ [X/H]   & $\Delta \tef$ & $\Delta$log $g$ & $\Delta v_t$      & $\Delta$[Fe/H]  & Param\tablefootmark{a} & Obs\tablefootmark{b} & Total\tablefootmark{c}  \\
{}       & LTE              &   \textbf{$\pm$}11 K       &  \textbf{$\pm$}0.03 dex      & \textbf{$\pm$}0.02 km\ s$^{-1}$ & \textbf{$\pm$}0.01 dex       &                        &                      &       \\
{}       & (dex)            & (dex)         & (dex)           & (dex)             & (dex)           & (dex)                  & (dex)                & dex   \\
\hline
C        &  0.059 & 0.006 & 0.008 & 0.001 & 0.000 & 0.010 & 0.004 & 0.011 \\
Na       &  0.069 & 0.005 & 0.001 & 0.000 & 0.000 & 0.005 & 0.010 & 0.011 \\
Mg       &  0.076 & 0.009 & 0.005 & 0.004 & 0.001 & 0.011 & 0.003 & 0.011 \\
Si       &  0.075 & 0.003 & 0.002 & 0.001 & 0.001 & 0.004 & 0.007 & 0.008 \\
Ca       &  0.069 & 0.007 & 0.004 & 0.000 & 0.004 & 0.009 & 0.007 & 0.011 \\
Sc       &  0.083 & 0.009 & 0.003 & 0.001 & 0.000 & 0.010 & 0.007 & 0.012 \\
Ti       &  0.076 & 0.010 & 0.001 & 0.003 & 0.000 & 0.010 & 0.006 & 0.012 \\
Cr       &  0.066 & 0.010 & 0.003 & 0.006 & 0.000 & 0.012 & 0.006 & 0.013 \\
Mn       &  0.073 & 0.007 & 0.001 & 0.003 & 0.001 & 0.008 & 0.001 & 0.008 \\
Fe       &  0.070 & 0.009 & 0.001 & 0.005 & 0.000 & 0.010 & 0.003 & 0.011 \\
Co       &  0.074 & 0.009 & 0.003 & 0.001 & 0.001 & 0.010 & 0.010 & 0.014 \\
Ni       &  0.085 & 0.006 & 0.000 & 0.003 & 0.001 & 0.007 & 0.003 & 0.007 \\
Cu       &  0.067 & 0.006 & 0.002 & 0.002 & 0.001 & 0.007 & 0.010 & 0.012 \\
Zn       &  0.066 & 0.002 & 0.000 & 0.007 & 0.003 & 0.008 & 0.007 & 0.011 \\
Y        &  0.086 & 0.001 & 0.011 & 0.009 & 0.003 & 0.015 & 0.006 & 0.016 \\
Zr       &  0.085 & 0.001 & 0.014 & 0.003 & 0.003 & 0.016 & 0.008 & 0.017 \\
Ce       &  0.084 & 0.003 & 0.015 & 0.002 & 0.003 & 0.016 & 0.007 & 0.017 \\
Nd       &  0.088 & 0.003 & 0.015 & 0.002 & 0.003 & 0.016 & 0.003 & 0.016 \\
C(CH)    &  0.062 & 0.009 & 0.000 & 0.001 & 0.008 & 0.012 & 0.010 & 0.016 \\
N(NH)    &  0.063 & 0.010 & 0.001 & 0.001 & 0.008 & 0.013 & 0.002 & 0.013 \\

\hline 

\end{tabular}
\tablefoot{\\
\tablefoottext{a}{Adding errors in stellar parameters} \\
\tablefoottext{b}{Observational errors} \\
\tablefoottext{c}{Total errors (stellar parameters and observational)}\\
}
\end{table*}

We developed a code\footnote{Our code is freely available online at \object{https://github.com/ramstojh/terra}} in python to make this calculation easier. The code first computes  the convective mass of a solar twin, and then estimates the mass of the rocky material missing in the convective zone. The convective mass is calculated making a double interpolation to the mass and metallicity of the solar twin. The convective masses adopted for this calculation are from \citet{Siess:2000}\footnote{\object{http://www.astro.ulb.ac.be/$\sim$siess}} . 

\begin{figure}
\centering
\includegraphics[width=9cm]{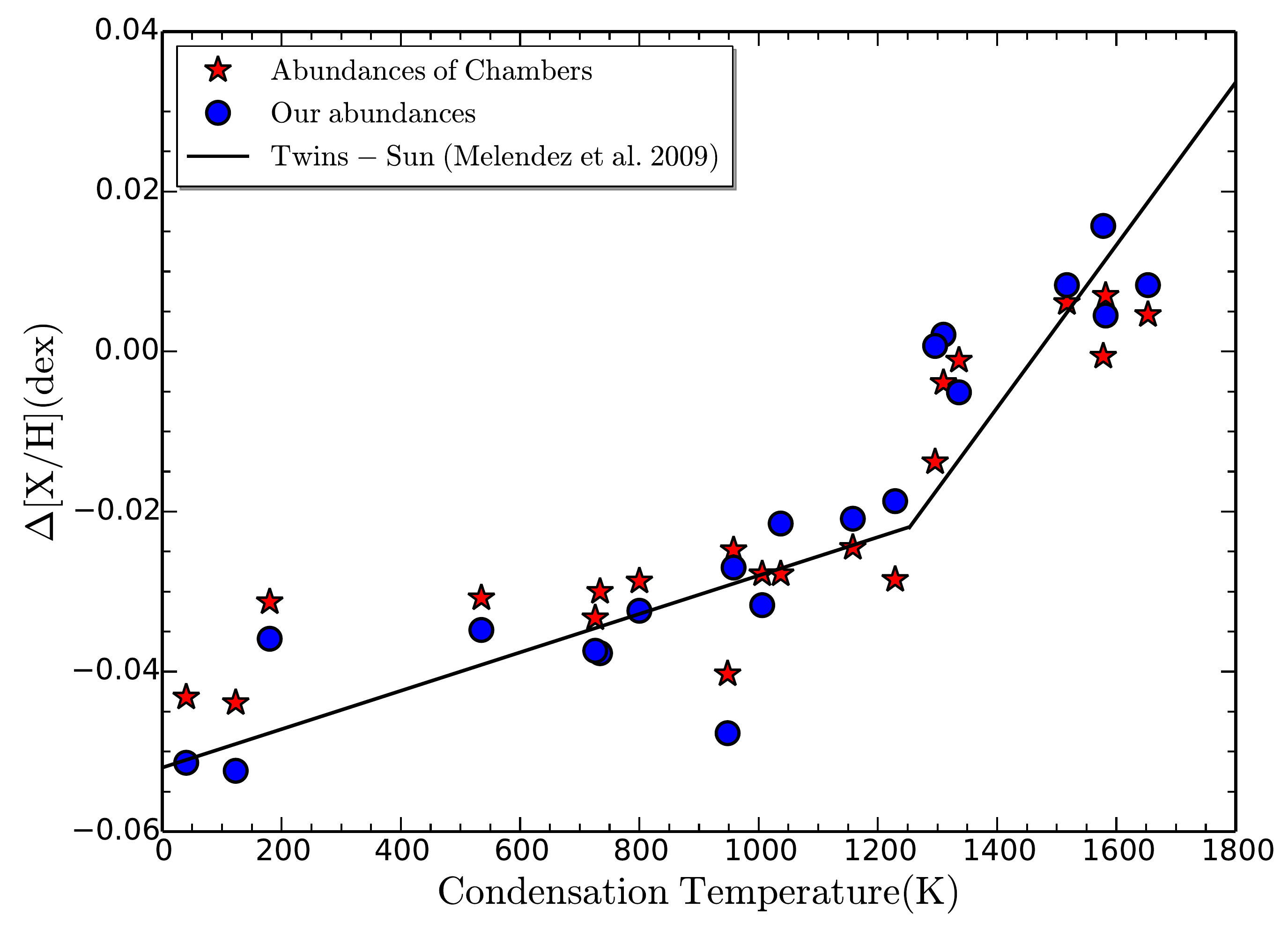}
\caption{Composition of the solar photosphere adding $4\ \rm{M}_{\oplus}$ in mixture of rocky material (Earth plus meteoritic) to the convective zone of the Sun.} 
\label{fig:comparing}
\end{figure}

Once the convective mass is obtained, the code computes the convective mass of each element present in the Sun, adopting the abundances of \cite{Asplund:2009}. Adopting the same approach, we use the abundances of \cite{Wasson:1988} and \cite{Allegre:2001} for the chemical composition of CM chondrites and the Earth. This is explained in more detail in appendix \ref{app:rocky} (see also \citet{Chambers:2010, Mack:2014}). All our calculations for Inti 1 are listed in Table \ref{tab:results}.

To verify our results, we reproduced the calculation of \cite{Chambers:2010} in Fig. 8, adding a mixture of $2\ \rm{M_{\oplus}}$ of Earth-like material and $2\ \rm{M_{\oplus}}$ of chondrite-like material. Our abundances are represented by blue filled circles, while the abundances of Chambers by filled red stars. Our results agree well with the calculation of Chambers, who used the same meteoritic composition as us, but  he adopted another reference for the Earth \citep{Waenke:1988}.


\section{Conclusions}
\label{sec:5}
Our atmospheric parameters computed relative to the Sun for HD 45184 are $T_{\rm{eff}} = 5864 \pm 9$ K, log $g = 4.45 \pm 0.03$ dex, $v_{t} = 1.11 \pm 0.02$ $\rm{km\ {s}}^{-1}$, and [Fe/H]$ = 0.04 \pm 0.01$ dex, and for the faint solar twin Inti 1 are $T_{\rm{eff}} = 5837 \pm 11$ K, log $g = 4.42 \pm 0.03$ dex, $v_{t} = 1.04 \pm 0.02$ $\rm{km\ {s}}^{-1}$, and [Fe/H]$ = 0.07 \pm 0.01$ dex, meaning that our star Inti 1 is a solar twin \citep{Ramirez:2009}. In Table \ref{tab:magnitude} we showed that the colors of Inti 1 are identical to the Sun within errors. Thus our faint solar solar twin and the distant solar twin found by \citet{Nascimento:2013} could be used to calibrate different photometric surveys.

The abundances of HD 45184 shows an excellent agreement with the highly precise work by  \citet{Nissen:2015}. HD 45184 and Inti 1 present enhancement in the refractory elements compared to volatiles species with an offset of 0.076 dex and 0.025 dex, respectively. This demonstrates that the same chemical anomalies found when the Sun is compared to local solar twins are also present when compared to distant solar twins (albeit the refractory enhancement of Inti 1 is smaller than for the nearby solar twin HD 45184). Using our code and theoretical models from \citet{Siess:2000}, we estimated the mass of rocky material present in the convective zone for HD 45184 ($\approx 3.5\ \rm{M_{\oplus}}$) and Inti 1 ($\approx 1.5\ \rm{M_{\oplus}}$), relative to the Sun.

The lithium abundances are usually determined from the 6708 \AA \ resonance line (or sometimes from the 6104 \AA \ subordinate line). Hence, they were not measured because of our limited spectral coverage (3190-5980 \AA). However, future observations will allow us to determine the Li abundance of Inti 1, and verify if the star follows the lithium-age correlation \citep[e.g.,][]{Monroe:2013}.

\begin{acknowledgements}
J.Y.G. acknowledges support by CNPq. J.M thanks support from FAPESP (2012/24392-2). We are grateful to the many people who have worked to make the Keck Telescope and its instruments a reality and to operate and maintain the Keck Observatory. The authors wish to extend special thanks to those of Hawaian ancestry on whose sacred mountain we are privileged to be guests. Without their generous hospitality, none of the observations presented herein would have been possible.
\end{acknowledgements}

\bibliographystyle{aa}
\bibliography{Objects}

\begin{appendix}
\label{app:rocky}
\section{Estimating the missing mass of rocky material}
Adopting the abundance of \cite{Asplund:2009} for the chemical composition of the Sun ($\rm{A_{sun}}$), and using [Fe/H] of a given star, we could estimate the mass of rocky material missing from its convective zone. First we have to compute the mass of each element, which is given by
\begin{equation}
\rm{X_{conv\_ mass}} = A \times 10 ^{(\mathrm{\rm{A_{sun}} \ + \ [Fe/H]})},
\end{equation}
where A is the atomic weight. Then the convective mass of each element is given by
\begin{equation}
Convective_{mass} = \rm{M_{conv} \times M_{\odot}} \times \frac{X_{conv\_ mass}}{\sum X_{conv\_ mass}},
\end{equation}
where $\rm{M_{conv}}$ is the Sun's convective mass ($\sim 0.02 \rm{M_{\odot}}$). For computing the meteoritic mass, we adopted the abundance of \cite{Wasson:1988} as composition of chondrites, and for the terrestrial mass we adopted the abundance of \cite{Allegre:2001},  following the equations:
\begin{eqnarray}
Meteoritic_{mass} = \rm{M_{\oplus} \times \frac{X_{met\_ mass}}{\sum X_{met\_ mass}}} \\
Terrestrial_{mass} = \rm{M_{\oplus} \times \frac{X_{terr\_ mass}}{\sum X_{terr\_ mass}}}, 
\end{eqnarray}
where $\rm{X_{met\_mass}}$ and $\rm{X_{terr\_mass}}$ are the masses of each element in Earth masses. 

Then the mass of a given element in the convective envelope, necessary to explain the observed abundance difference, is given by 
\begin{equation}
\Delta M = \log \left(1+\frac{M_{rock}}{Convective_{mass}} \right),
\end{equation}
where $M_{rock}$ is the mass of the missing rock, and is given by $M_{rock} = \alpha \times Meteoritical_{mass} + \beta \times Terrestrial_{mass}$, where $\alpha$ and $\beta$ are given in units of Earth masses.\\

\end{appendix}

\end{document}